\documentclass{MolSpaLab}
\usepackage{graphicx}

\begin{document}
\TitreGlobal{RSC Glasgow 2009}
\title{Simple hydrogen-bearing molecules in translucent molecular clouds}
\author{\FirstName T. \LastName Weselak}
\address{Department of Physics, Kazimierz Wielki University,  Weyssenhoffa 11,
85-072 Bydgoszcz, Poland}
\author{\FirstName J. \LastName Kre{\l}owski}
\address{Center for Astronomy, Nicolaus Copernicus
University,
Gagarina 11, Pl-87-100 Toru{\'n}, Poland
}

\runningtitle{Simple hydrogen-bearing molecules in translucent molecular clouds}
\setcounter{page}{1}

\maketitle
\begin{abstract}
We demonstrate relations between column densities of simple molecules: 
CH, CH$^{+}$, H$_{2}$ and OH. The H$_{2}$, CH and OH molecules seem to 
occupy the same environments because of tight relations between their 
column densities. 
In contrary to this CH$^{+}$ column density does not correlate with 
those of other simple molecules. 
\end{abstract}

\section{Introduction}

Molecular hydrogen H$_{2}$ is the most abundant molecule in the
interstellar medium (ISM) with column densities exceeding
10$^{19}$ cm$^{-2}$ toward examined OB stars; its ultraviolet
spectrum may be detected only by spaceborn instruments. Spectra of
translucent clouds are also characterized by features of other
simple molecules such as CH, CH$^{+}$, OH, NH, and CN available to
ground-based telescopes.\\
Recently it was shown by  Weselak et al. (2008a) that column
densities of H$_{2}$ and CH$^{+}$ show large scatter suggesting no
relation between these two molecules. On the other hand CH
molecule is closely related to molecular hydrogen (Mattila 1986,
Weselak et al. 2004). Also the relation between column densities
of the CH and OH molecules is very good (Weselak et al. 2009).\\
Here we present relations between column densities of simple
hydrogen-bearing molecules (H$_{2}$, CH, CH$^{+}$, OH). The
observational material is based on high resolution and high S/N
ratio observations using five echelle spectrographs. Column
densities of the H$_{2}$ molecule were obtained from the
literature.

\section{The Observational Material}
\label{section1}
Our observing material  was acquired using five echelle spectrometers:
\begin{itemize}

  \item MAESTRO fed by the 2-m telescope of the Observatory 
at Peak Terskol (see http://www.terskol.com\ /telescopes/3-camera.htm)
  \item Feros spectrograph, fed with the 2.2m ESO 
telescope in Chile\\ (see http://www.ls.eso.org/lasilla/sciops/2p2/E2p2M/FEROS/)
  \item fiber-fed echelle spectrograph installed at 
1.8-m telescope of the Bohyunsan Optical Astronomy Observatory (BOAO) in South Korea
  \item HARPS spectrometer, fed with the 3.6m ESO 
telescope in Chile\\ (see 
http://www.ls.eso.org/lasilla/sciops/3p6/harps/)
  \item UVES spectrograph at Paranal in Chile. 
For more information see:\\ 
http://www.eso.org/sci/facilities/paranal/instruments/uves
\end{itemize}

To obtain column densities we used the relation of {Herbig (1968)
which gives proper column densities when the
observed lines are unsaturated:
\begin{equation}
N = 1.13\times10^{20} W_{\lambda}/(\lambda^{2} f),
\end{equation}
\noindent where W$_{\lambda}$ and $\lambda$ are in \AA\ and column
density in cm$^{-2}$. To obtain column density we adopted f-values
listed in Table 1.

\begin{figure*}[ht!]
\includegraphics[width=6.5 cm]{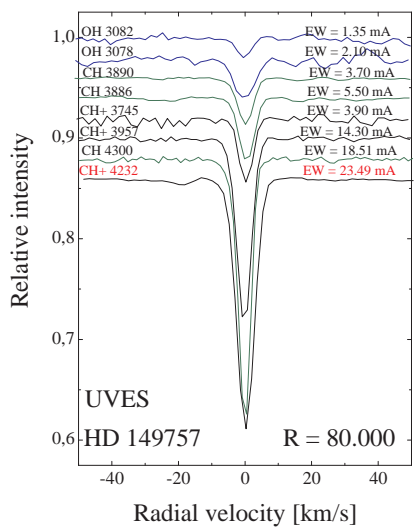}
\caption{  
Interstellar features
of the OH, CH and CH$^{+}$ molecules
(presented in Table 1) in the spectrum
of HD 149757. The CH$^{+}$ line at 4232 \AA\
is saturated. In this case to obtain  column
density we used the unsaturated CH$^{+}$ line 
at 3957 \AA.}

\end{figure*}

\newpage
\noindent
{\small{
{\bf{Table 1}}: Adopted molecular parameters. References: 1~--~Weselak et al. (2009b), 2~--~Felenbok and Roueff (1996), 
3~--~Gredel et al. (1993), 4~--~Weselak et al. (2009a), 5~--~van Dishoeck and Black (1986)

\begin{tabular}{lrclclcccc}
\hline\hline
& & & & & & & & &\\
Species &   Vibronic band               &   Rotational lines        &   Position   &Ref.   &   f-value &Ref.    \\
& & &[\AA] &  & & & & &\\
\hline
& & & & & & & & &\\
OH  &   A$^{2}\Sigma^{+}$ -- X$^{2}\Pi_{i}$ (0, 0)  	&   Q$_{1}$(3/2)+$^{Q}P_{21}$(3/2)&   3078.443    &1  &   0.00105 &2  \\
    &                   		(0, 0)  	&   P$_{1}$(3/2)            	  &   3081.6645   &1  &   0.000648 &2  \\
CH  &   A$^{2}\Delta$ -- X$^{2}\Pi$ (0, 0)		&   R$_{2e}$(1) + R$_{2f}$(1)     &   4300.3132   &3  &   0.00506  &5 \\
    &   B$^{2}\Sigma^{-}$ -- X$^{2}\Pi$ (0, 0)  	&   Q$_{2}$(1)+$^{Q}$R$_{12}$(1)  &   3886.409    &3  &   0.00320 &3  \\
    &                  			 (0, 0)  	&   $^{P}$Q$_{12}$(1)          	  &   3890.217    &3  &   0.00210 &3  \\
CH$^{+}$&   A$^{1}\Pi$ -- X$^{1}\Sigma^{+}$ (0, 0)   	& R(0)               		  &   4232.548    &3  &   0.00545 &3  \\
    &   				(1, 0)          &   R(0)                	  &   3957.689    &4  &   0.00342 &4  \\
    &  					 (2, 0)         &   R(0)               		  &   3745.308    &4  &   0.00172 &4  \\

\hline
\end{tabular}
}}

\section{Results}
\label{section2}

\begin{enumerate}
\item{
The CH cation seem to be formed in
other reaction pathways since their column densities do not correlate
with those of H$_{2}$ and OH (Figs 2a and 2b).}

\item{
The column densities of both OH, CH and H$_{2}$ molecules are closely related as
seen in Figs 3a and 3b. This relation suggests that both species
originate and are preserved in the same environments.
One atypical object, HD 34078, misses the relation between column densities of 
OH and CH molecules and also between the column densities of CH and H$_{2}$
molecules.}

\item{
Interstellar molecules such as H$_{2}$, CH and OH seem to be
closely connected in translucent clouds. 
The CH molecule may be used as an OH and H$_{2}$ tracer, as its
features are easily accessible and well--related to the
abundances of hydrogen and hydroxyl.
}
\end{enumerate}

\begin{figure*}[ht!]

\includegraphics[width=6.5 cm]{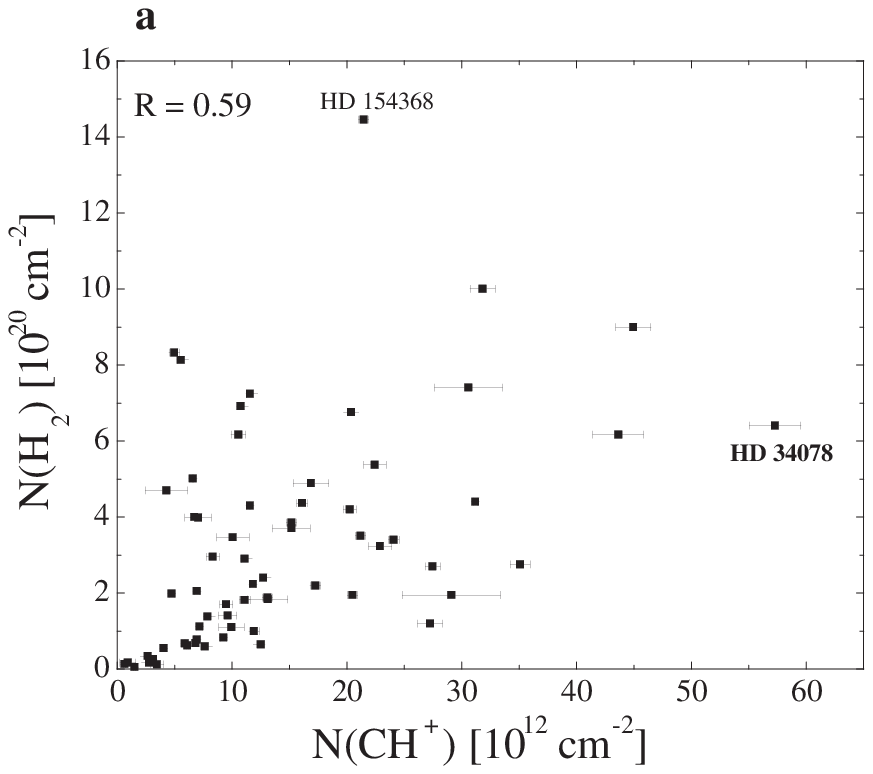}
\includegraphics[width=6.7 cm]{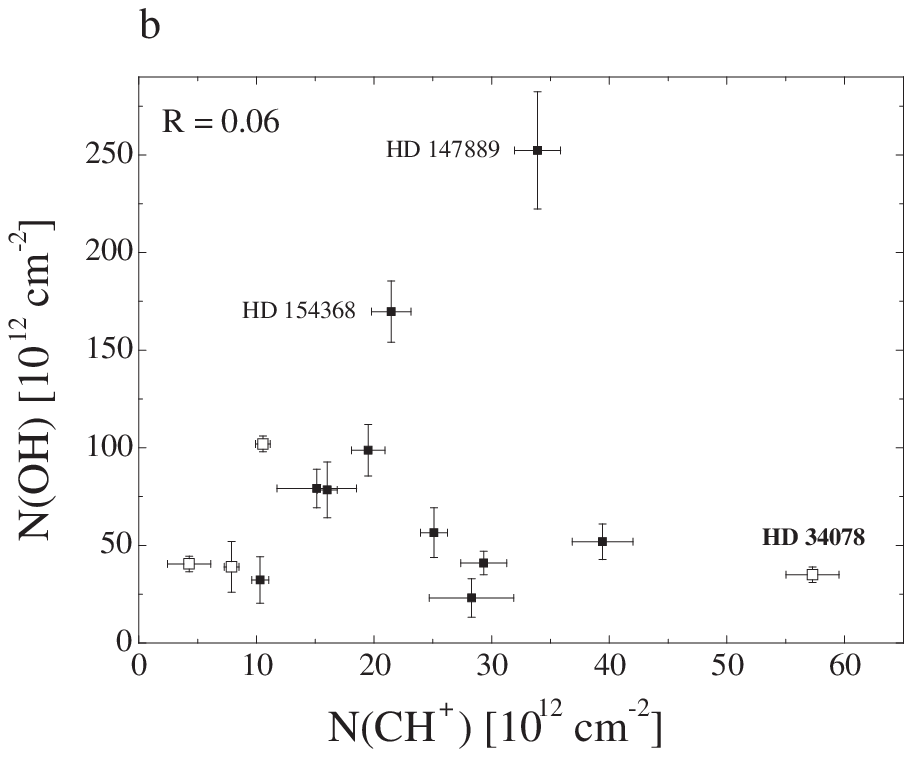}
\caption{ The poor relation 
between column densities of CH$^{+}$
and H$_{2}$ molecules (a) and 
also between the column densities of 
OH and CH$^{+}$  (b). Correlation coefficients
are presented at the top-left in each case. 
}
\label{fig}
\end{figure*}

\begin{figure*}[ht!]
\includegraphics[width=6.5 cm]{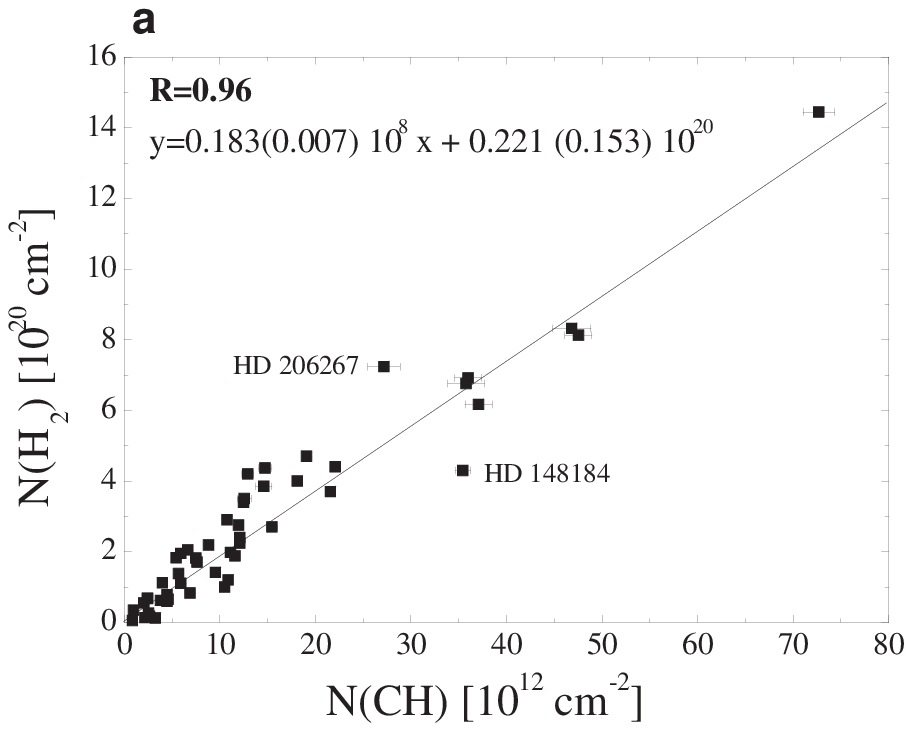}
\includegraphics[width=6.5 cm]{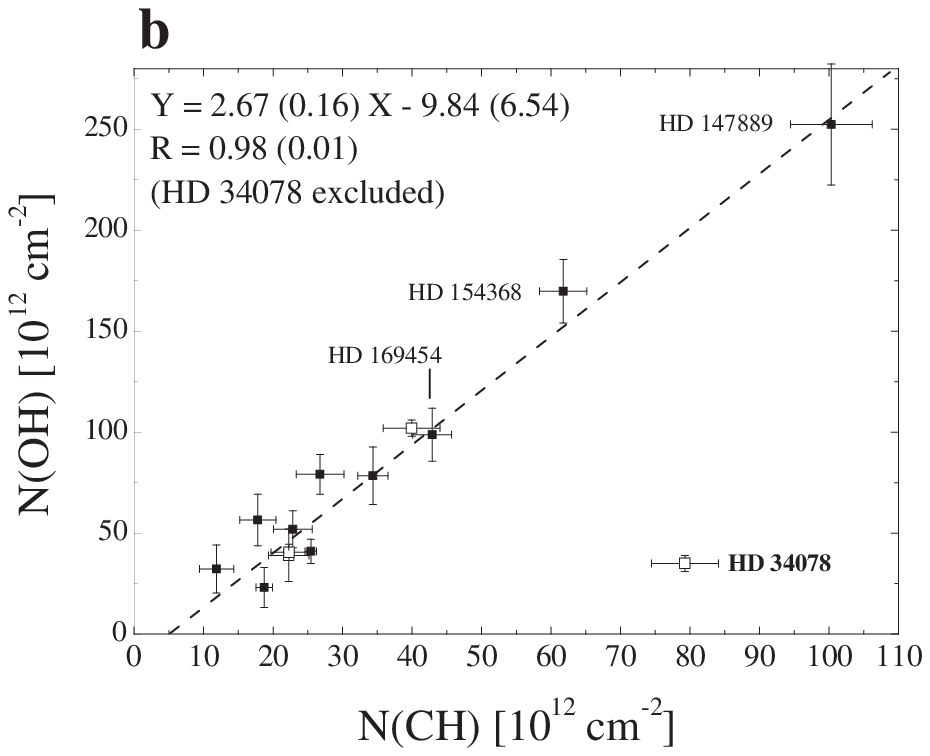}
\caption{ 
The very good relation 
between column densities of
H$_{2}$ and CH molecules (a) and 
also between the column densities of 
OH and CH (b). With the dashed line
we present the relation with one
data-point excluded (HD 34078).}
\label{fig}
\end{figure*}

\newpage
\section{Acknowledgements}
\label{section3}
Authors acknowledge the financial support: JK and TW
acknowledge that of the Polish State during the period
2007 - 2010 (grant N203 012 32/1550).

\end{document}